\documentstyle[preprint,aps]{revtex}
\newcommand{\beq}{\begin{equation}}
\newcommand{\eeq}{\end{equation}}
\newcommand{\bdis}{\begin{displaymath}}
\newcommand{\edis}{\end{displaymath}}
\newcommand{\bea}{\begin{eqnarray}}
\newcommand{\eea}{\end{eqnarray}}
\newcommand{\barr}{\begin{array}}
\newcommand{\earr}{\end{array}}

\begin{document}
\bibliographystyle{prsty}
%
\title{Laplacian Fractal Growth in Media with Quenched Disorder}
\author{R.Cafiero$^1$, A. Gabrielli$^{1,2}$,  M. Marsili$^3$, L. 
Pietronero$^1$ and L. Torosantucci$^1$ }
\address{
$^1$Dipartimento di Fisica, Universit\'a di Roma "La Sapienza",
P.le Aldo Moro 2,I-00185 Roma, Italy; and INFM, unit\'a di Roma I}
\address{
$^2$Dipartimento di Fisica, Universit\'a di Roma "Tor Vergata", 
Via della Ricerca Scientifica 1, I-00133 Roma}
\address{
$^3$Institut de Physique Th\'eorique, Universit\'e de Fribourg, 
P\'erolles CH-1700 Fribourg, Suisse}

\date{\today}
\maketitle

\begin{abstract}
We analyze the combined effect of a Laplacian field and quenched 
disorder for the generation of fractal structures with a study, both 
numerical and theoretical, of the quenched dielectric breakdown 
model (QDBM). The growth dynamics is shown to evolve from the 
avalanches of invasion percolation (IP) to the smooth growth of 
Laplacian fractals, i. e. diffusion limited aggregation (DLA) and the 
dielectric breakdown model (DBM). The fractal dimension is strongly 
reduced with respect to both DBM and IP, due to the combined effect 
of memory and field screening. This implies a specific relation 
between the fractal dimension of the breakdown structures (dielectric 
or mechanical) and the microscopic properties of disordered 
materials.
\end{abstract}

\smallskip
\newpage

The growth of fractal structures is usually described in terms of 
physical models characterized by an irreversible dynamics and a 
degree of self-organization. These models can be divided in two 
broad classes. In the first one the dynamics is {\em stochastic} (
modulated by a field), and its prominent examples are diffusion 
limited aggregation (DLA) \cite{dla}, the dielectric breakdown 
model (DBM) \cite{dbm} and their variants. The second class is 
characterized by a deterministic dynamics in a {\em quenched} 
random medium. An example is Invasion Percolation (IP) 
\cite{Wilk} and a related model is the Bak and Sneppen (BS) 
\cite{bak1} model of biological evolution. Concerning the growth 
dynamics, some of these models grow in a smooth way (i.e. DLA and 
DBM), while other ones grow by {\em avalanches}  (i.e. IP) 
\cite{bak1}. 

Our understanding of the origin of fractal structures for the two 
classes is based on the idea that an {\em effective screening} is 
present in the scale invariant dynamics \cite{Raf}. However, the 
origin and the properties of this screening are very different. For 
Laplacian fractals it arises from the geometrical screening of the 
electric field around the structure. For the case of quenched 
dynamics, instead, there is no field and an effective screening 
develops as a memory effect from the quenched dynamics itself. 

The two kinds of dynamical models, stochastic with a field and 
quenched without a field, represent however two extreme limiting
 cases of real natural phenomena which 
usually present both features. For 
example dielectric breakdown and fracture 
propagation in disordered solids represent 
important cases in which these 
elements are combined \cite{qdbm,qfrac}. 

In this letter we address the question of 
understanding how the combination of these two basic elements 
operates and leads to new phenomena. From a theoretical point of 
view this requires the unification of the concepts developed for the 
two limiting cases. Together with suitable simulations, this allows us 
to link the microscopic properties of disordered materials to the 
resulting fractal structures.

In the original DBM the local field $E_i$ across a perimeter bond $i$ 
of the structure is related to the probability of growth by $p_i 
\propto E_i^{\eta}$, where $\eta$ is a parameter characterizing the 
strength of the link between the field and the growth 
probability. The growth probability of a bond is then normalized 
with that of all the other perimeter bonds. If we consider 
the generalization of such a process in a random medium 
\cite{qdbm,qfrac} each bond can 
be associated to a quenched variable $x_i \in [0,1]$, extracted from 
a given probability distribution, which defines the strength of the 
bond (resistivity) with respect to the breakdown. The growth 
process proceeds by breaking the weakest bond. Without the field 
this means to break the bond with $\min_i \{ x_i \}$; this is 
Invasion Percolation \cite{Wilk}. If we introduce the local field 
$E_i$, the tendency to break will be modulated by the field itself. 
This requires the introduction of a 
new variable:
\begin{equation}
y_i(t)=\frac{x_i}{\left[E_i(t)\right]^{\eta}}
\label{vary}
\end{equation}
which takes into account this effect. Growth at time $t$ occurs at 
the bond with the smallest $y_i(t)$.

The probability density of $x_i$ is usually assumed of a power 
law form \cite{qdbm}
\begin{equation}
p_0(x)=a x^{a-1}
\label{densx}
\end{equation}
where the parameter $a$ ($a \in [0, \infty]$) models the strength of 
the material. Small values of $a$ ($a<<1$) correspond to a fragile 
material in which most bonds are easy to break. Large values of $a$ 
($a>>1$) correspond instead to a strong material.
The limit $a \to \infty$ ($\eta \to \infty$) implies that growth 
occurs deterministically at the bond with the largest field. The case 
$\eta=0$ eliminates the effect of the field and leads to the IP model. 
On the other hand, there is no obvious limit that brings us back to 
the original (stochastic) DBM. In this respect the present model, that 
we may call Quenched DBM (QDBM), contains the elements of both 
the DBM and IP models, but it has a simple limiting case only 
towards IP. 

The effect of the field on the quenched variables is to modify the 
extension of its distribution for a given bond, as shown in fig. 
\ref{fig1}. So, while the extension of $x_i$ is on the $[0,1]$ interval 
for all bonds, the distribution of $y_i$ will have an extension 
depending on the local field $E_i$ (fig. \ref{fig1}). It is convenient to 
use a normalized field (at time $t$)
\begin{equation}
\left[ E^{'}_i(t) \right]^{\eta}=\left[ \frac{E_i(t)}{E_{max}(t)} 
\right]^{\eta}
\label{normel}
\end{equation}
where $E_{max}(t)$ is the maximum value of the field among the 
perimeter bonds at time $t$. Using this field we have 
$y_i(t)=x_i/[E'_i(t)]^{\eta}$, with a probability density (from eq. 
\ref{densx}:
\begin{equation}
p_0(y)=a[E'_i(t)]^{a \eta} y^{a-1}
\label{ydensin}
\end{equation}
and therefore the range of the variable $y_i(t)$ for the bond with 
the largest field is always $0 \leq y_i(t) \leq 1$, which is more 
convenient for the analysis.

A fundamental characterization of the growth process comes from the 
acceptation profiles, corresponding to the distributions of the $x_i$, 
$y_i$ for the grown bonds. In fig. \ref{fig2} we 
report, for parameter values $a=1$ and $\eta=1$, the 
distribution of the variables $x_i$ of the grown bonds, 
while the insert refers to the analogous distribution for the $y_i$.

The behaviour of these distributions as a function of the total growth 
time $t$ shows the {\em self-organization} of the system towards specific limiting distributions.
The distribution of the $x_i$ converges to a function which extends 
on the entire $[0,1]$ interval. This implies that a value of $x_i$ close 
to $1$ (unfavourable) may actually prevail on all others because of 
the effect of the field. This situation is very different from the case 
of IP, in which the analogous distribution is a theta function with 
a discontinuity at $x=p_c<1$, where $p_c$ is the critical 
bond percolation probability. The existence of such a threshold implies 
that the dynamics evolves by scale invariant avalanches 
\cite{bak1,rtslung}.

The acceptation profile of the $y_i$ (insert of fig. \ref{fig2}) 
seems to converge to a theta function, but its meaning is quite 
different than for the case of IP. In fact, the distribution must be 
zero for $y>1$ by construction, because at least the variable with 
the maximum field must be smaller than one. Therefore the bonds 
that are allowed to grow must refer to values of $y$ in $[0,1]$. On 
the other hand its approximately flat behaviour is quite non 
trivial and it indicates that the effect of the field 
dominates with respect to that of 
the quenched disorder. The best candidates for growth are near the 
tips because of the field, so that memory effects are limited with 
respect to IP and the process is close to deterministic DBM (which 
would give $D=1$).

The fractal structures corresponding to QDBM are shown in 
fig.\ref{fig3}, while in Table \ref{table} we report the values of the 
QDBM fractal dimensions for different values of $a$ and $\eta$. 
These values give a strong indication for the invariance property
\begin{equation}
D_f(\eta;a)=D_f(\eta \cdot a)
\label{dfsimm}
\end{equation}
This result is quite interesting and non trivial in view of the very 
different role that the parameters $\eta$ and $a$ play in the 
growth process. It may be explained by observing that if we make the transformation 
\beq x_i \to x_i^{a},\,\,\eta \to \eta'=a/ \eta,
\label{trans}
\eeq
the ordering of variables $y_i$ do not change and the dynamical evolution leads to the same fractal cluster.

From the theoretical side the growth process is deterministic and it 
arises from the quenched disorder modulated by the field of the 
structure itself. A suitable approach to deal with a problem of fractal 
growth with quenched disorder is the method of the Run Time 
Statistics (RTS) \cite{46matt,rts,rtslung}. 
It consists in the mapping of a 
quenched dynamics into a stochastic one with {\em cognitive 
memory}. 
The basic concept of the mapping is the following: if a bond has 
"lost" many times, there is a finite probability that it will never 
grow, even after an infinite time. This introduces an effective 
dynamical screening which is at the basis of fractal properties. The 
RTS method provides a systematic technique to describe this 
phenomenon, which is typical of models with quenched disorder. 

In the present case the situation is more complex in view of the 
modulation induced by the field. To this purpose it is convenient to 
introduce the following concepts. To each interface bond we assign 
an effective probability density $p_{\tau,t,i}(y)$, where $i$ gives the 
position of the bond, $t$ is the total growth time and $\tau$ is the 
time during which the bond has been part of the growth interface 
without being selected ("age"). As soon as a bond becomes part of 
the growing interface ($\tau=0$), its distribution is the original one
$p_{0,t,i}(y)=p_0(y)=a[E'_i(t)]^{a \eta} y^{a-1}$. After a time $\tau$ 
the density is modified in a way that is conditional to the growth 
history of the bond. Using the rules of the conditional and composed 
probabilities \cite{feller} and by performing an 
average over quenched disorder, 
we can derive an expression for the growth 
probability $\mu_{\tau,t,i}$ 
of the variable $i$ at time $t$. This implies that $i$ is the extremal 
variable and the expression of its growth probability is 
\cite{qdbmrtslun}
\begin{equation}
\mu_{\tau,t,i}= \int_{0}^{1}
dy \: p_{ \tau ,t,i} (y) \prod_{m}
\int_{ y}^{Y_m} dy_{m} \: p_{ \tau_m ,t,m} (y_{m}),
\label{muMQM}
\end{equation}
where $Y_m=\frac{1}{[E'_{m}(t)]^{\eta}}$ and 
the product accounts for the competition between the 
growing bond and the other perimeter bonds. In the same way we 
can obtain an equation for the 
update from time $t$ to $t+1$ of the effective densities:
$$p_{\tau+1,t+1,i}(y) = \left[r_i
p_{\tau,t,i}(r_i y) 
\int_{0}^{r_i y}dy_{j}\: 
p_{\tau_j,t,j}(y_j)
\theta(\frac{1}{E'_{j}(t)^{\eta}} - y_{j}) \right.$$
\begin{equation}
\left. \prod_{m\!\neq\!j,i} \theta(\frac{1}{E'_{m}(t)^{\eta}}-y_j) \cdot 
\int_{y_{j}}^{\frac{1}{E'_{m}(t)^{\eta}}}dy_{m}\: p_{\tau_m,t,m}
(y_{m}) \right] \frac{1}{\mu_{\tau,t,i}}  
\label{agg}
\end{equation}
where $r_i= \left(\frac{E'_{i}(t+1)}{E'_{i}(t)} 
\right)^{\eta}$ and we accounted for the change of 
the electric field of the 
variable $i$ from time $t$ to time $t+1$.

Equations \ref{muMQM}, \ref{agg} provide the mapping of the 
original quenched problem into a stochastic one defined by growth 
probabilities. The specific realizations of this stochastic process have 
the same statistical weight as those of the original quenched 
problem. The approximation in this mapping consists in neglecting 
the geometrical correlations between probability densities of different 
bonds, 
but this can be shown to be exact in the limit of large systems 
\cite{correl}. Other approximations of technical nature will be 
necessary for the practical use of this mapping in the FST scheme for 
the calculation of the fractal dimension. 

From 
the expression \ref{muMQM} it is possible to derive the invariance 
property \ref{dfsimm} that we inferred from the simulations. In fact 
one can show that eq. \ref{muMQM} is invariant under the transformation
 \ref{trans} \cite{qdbmrtslun}.
We can also study the behaviour of screening effects in QDBM 
dynamics. An analysis of eqs. \ref{muMQM} and \ref{agg} gives 
\cite{qdbmrtslun}:
\begin{equation}
\mu_{\tau,t,i} \propto [E'_i(t)]^{a \eta (\tau+1)}
\label{scre1}
\end{equation}
The electric field of the bond $i$ at time $t$, expressed in terms of the 
initial value of the field $E'_i(t_0)$ ($t_0$ being the time at which the 
bond becomes part of the interface) and of the "age" $\tau$ 
of the bond is, in analogy with DBM \cite{mars}:
\begin{equation}
E'_i(t=t_0+\tau)=E'_i(t_0) e^{-c_{i,\tau} \tau}
\label{scrfiel}
\end{equation}
where $c_{i,\tau}$ is a function which depends on the growth 
history. By using eq. \ref{scrfiel} into eq. \ref{scre1} 
one obtains:
\begin{equation}
\mu_{\tau,t,i} \propto e^{-c_{i,\tau} a \eta \tau (\tau+1)}
\label{scrfin}
\end{equation}
This result shows that screening effects in QDBM 
are {\em stronger than in stochastic 
DBM} (where $\mu_{\tau,t,i} \propto e^{-c_{\tau,i} \tau}$). This 
explains therefore why QDBM has a fractal 
dimension much smaller than 
the usual DBM or DLA. Moreover, the exponential 
screening of the GPD of QDBM, 
compared to the power law screening ($\mu_{\tau,t,i} \propto 
\frac{1}{(\tau+1)^{\alpha}}$ \cite{rts}) corresponding 
to IP and similar models, 
 gives a further evidence for the absence of scale invariant 
avalanches in QDBM. In fact, as discussed also in 
\cite{rtslung}, the shape of the GPD of extremal processes 
determines the shape of the correspondent avalanche size 
distribution. Therefore, a power law distribution for the avalanche 
sizes can only arise from a power law screening of the growth 
probability of the interface bonds with respect to the growth time.

Having characterized the quenched dynamics 
of the QDBM in terms of 
RTS growth probabilities, we can now consider the use of the Fixed 
Scale Transformation (FST) \cite{fst2} method to analyze the fractal 
properties of the resulting structures. Such a step, however, is 
nontrivial and it requires the following considerations. The original 
dynamics is quenched as in IP, but in the end it does not lead to a 
critical threshold for the effective probability 
distribution. This implies 
that we cannot consider the growth by avalanches as in IP, which 
would correspond to a specific implementation of the FST 
\cite{rtslung}. On the contrary, the gaussian screening of eq. 
\ref{scrfin} implies that a fast convergence 
of the FST matrix elements 
will be achieved using the same scheme as 
in DLA and DBM \cite{fst2}.

A nontrivial problem, instead, is the identification of the {\em scale 
invariant dynamics} for this problem. In IP the 
simple extremal rules 
and the avalanche dynamics can be used to
 argue that the small scale 
dynamics is already scale invariant \cite{rtslung}. In DLA (DBM) the 
situation is more complex and it requires a specific renormalization 
study which also showed, however, that the small scale dynamics is 
rather close to the scale invariant one \cite{Raf}. In 
the present case a systematic 
study of the scale invariant dynamics would require the combined 
renormalization of the effective probability densities and of the 
electric fields of the bonds. At the moment, this appears far too 
complex. So, inspired by the results of the 
two cases we can treat, we 
assume that the small scale dynamics is a reasonable 
approximation to the scale invariant one.

In this way we can proceed to the explicit calculation
 of the FST matrix elements and of the corresponding 
fractal dimensions. The results are 
reported in table \ref{table1} for different values of 
$\eta$ with $a=1$ (the calculation is up to the second 
order in the FST scheme). 

These results allow us to understand the strong reduction of the 
fractal dimension for QDBM ($D=1.37$ for $\eta=1$) with respect to 
both IP ($D=1.8879$ \cite{rtslung}) and DBM 
($D=1.6406$ for $\eta=1$ 
and cilinder geometry \cite{fst2}). On the other hand, the values of 
$D$ we obtain from theory appear to be 
somewhat larger than those of the 
simulations. One can speculate that this may be due to the 
approximation used for the scale invariant dynamics. 

We would like to thank G. Caldarelli for useful discussions.

\begin{figure}
\protect\caption{Schematic picture of the 
role of a field ($E_i$) on the 
dynamics of QDBM. The two bonds have 
the same (flat) distribution for 
the variable $x_i$, which represents the 
microscopic properties of the 
material. However, once the bonds are subjected to different fields 
($E_1>E_2$), the effective variables for 
the breakdown process become 
the $y_i=x_i/E_i$, whose distributions are modulated by the field.}
 \label{fig1}
 \end{figure}
q
\begin{figure}
\protect\caption{Time evolution of the 
acceptation profiles for the variables 
$x_i$ and $y_i$ (insert) for the bonds which grow. For $t \to \infty$ 
the convergence towards limiting distributions corresponds to the self-
organized nature of the dynamics. The absence of a threshold value for 
$a(x)$ implies a smooth growth (not by avalanches).}
 \label{fig2}
 \end{figure}

\begin{figure}
\protect\caption{Examples of QDBM clusters for 
different values of $a \cdot 
\eta$. Note the small values of $D$ with respect to IP and DBM.}
\label{fig3}
 \end{figure}

\begin{table}
\begin{centering}
\begin{tabular}{cccc}
\hline
$\eta$ & $D(\eta;a=1)$ & $a$ & $D(a;\eta=1)$\\
\hline
$0.2$ & $1.33\pm0.02$ & $0.2$ & $1.35\pm0.02$\\
$0.5$ & $1.21\pm0.02$ & $0.5$ & $1.22\pm0.02$\\
$1.0$ & $1.15\pm0.02$ & $1.0$ & $1.15\pm0.02$\\
$2.0$ & $1.07\pm0.01$ & $2.0$ & $1.06\pm0.01$\\
$3.0$ & $1.02\pm0.01$ & $3.0$ & $1.03\pm0.01$\\
\hline
\end{tabular}
\protect\caption{Fractal dimension of QDBM clusters of size 
$512 \times 2048$, in cilinder geometry, for different values 
of the parameters $a$ and $\eta$.}
\label{table}
\end{centering}
\end{table}

\begin{table}
\begin{centering}
\begin{tabular}{cc}\hline
$\eta$ & $D(\eta)$\\
\hline\hline
$0.2$ & $1.70$\\
$0.5$ & $1.56$\\
$1.0$ & $1.37$\\
$2.0$ & $1.23$\\
$3.0$ & $1.15$\\
\hline
\end{tabular}
\protect\caption{Second order FST computation of the fractal 
dimension of QDBM, for different values of $\eta$ with $a=1$.}
\label{table1}
\end{centering}
\end{table} 

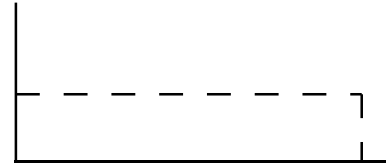
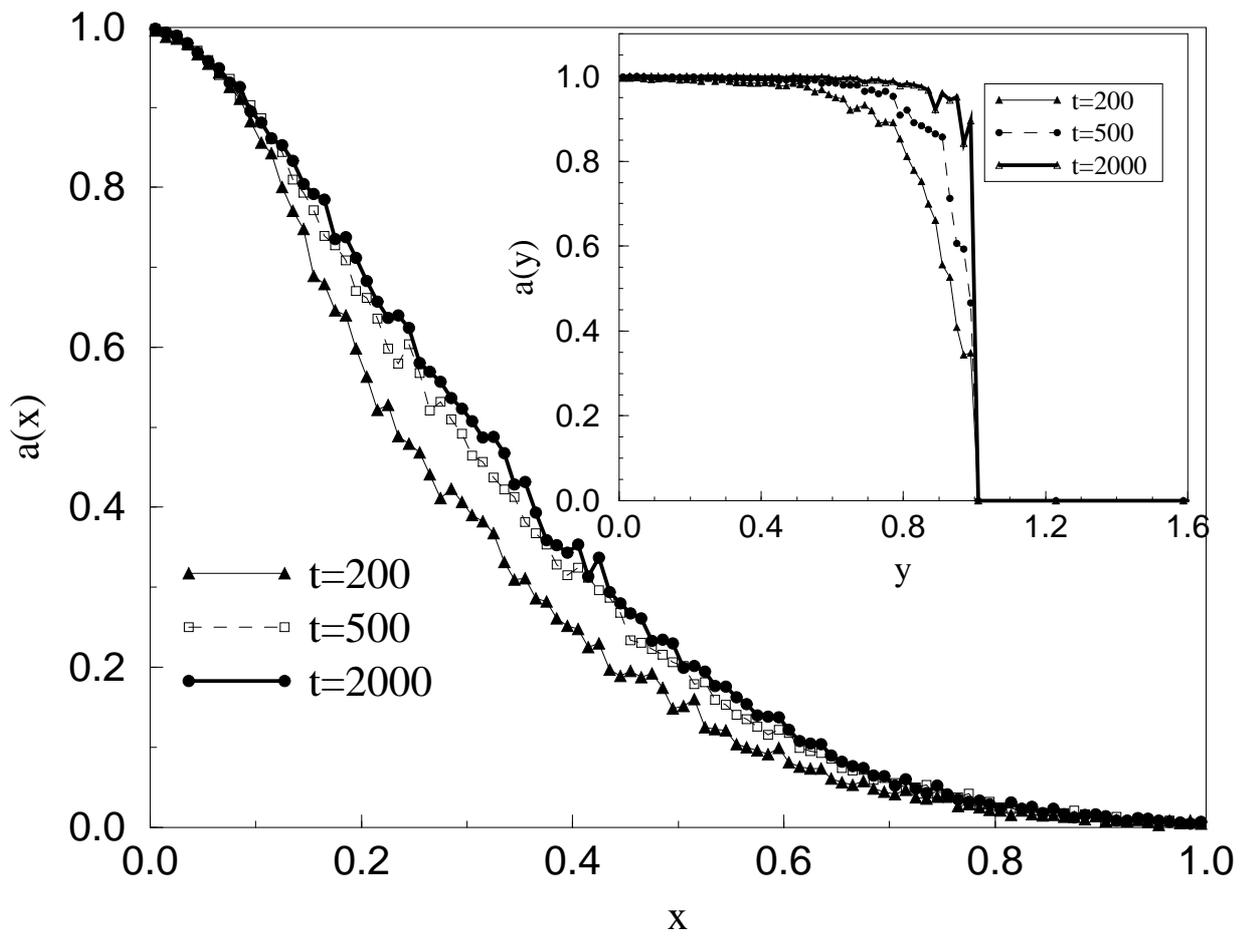
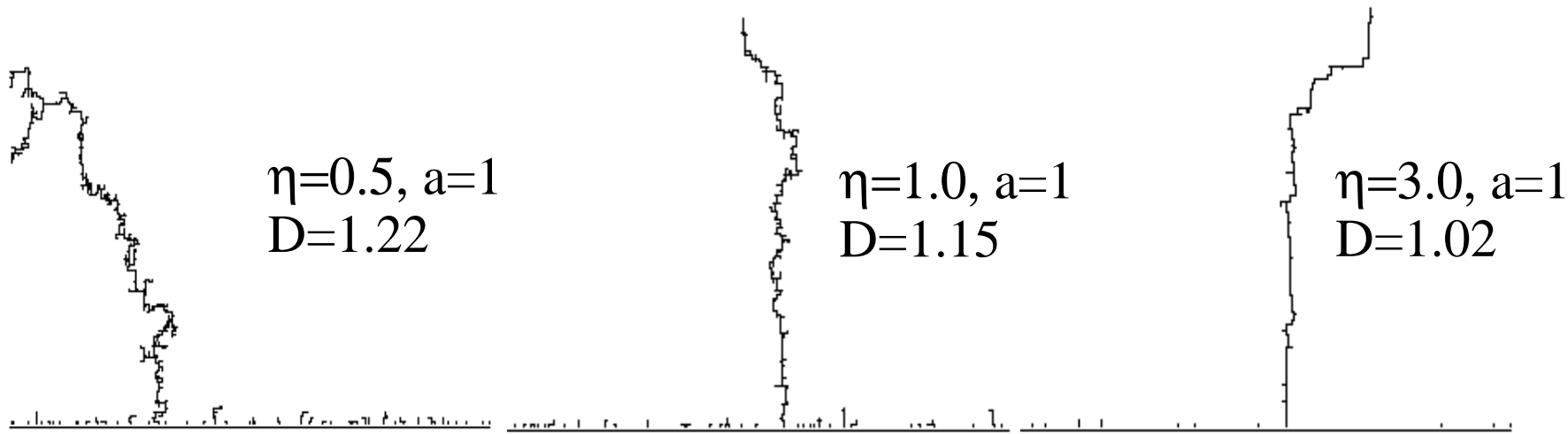
\end{document}